\documentclass[aps,prb,twocolumn,superscriptaddress]{revtex4}

\usepackage{graphicx}
\usepackage{amsmath}
\usepackage{amssymb}
\usepackage{braket}

\begin{document}

\title{Out-of-Time-Order Correlation for Many-Body Localization}
\author{Ruihua Fan}
\thanks{They contribute equally to this work. }
\affiliation{Institute for Advanced Study, Tsinghua University, Beijing, 100084, China}
\affiliation{Department of Physics, Peking University, Beijing, 100871, China }
\author{Pengfei Zhang}
\thanks{They contribute equally to this work. }
\affiliation{Institute for Advanced Study, Tsinghua University, Beijing, 100084, China}
\author{Huitao Shen}
\affiliation{Department of physics, Massachusetts Institute of Technology, Cambridge, MA 02139, USA}
\author{Hui Zhai}
\thanks{Electronic address: hzhai@tsinghua.edu.cn }
\affiliation{Institute for Advanced Study, Tsinghua University, Beijing, 100084, China}
\affiliation{Collaborative Innovation Center of Quantum Matter, Beijing, 100084, China}
\date{\today}

\date{\today }

\begin{abstract}

In this paper we first compute the out-of-time-order correlators (OTOC) for both a phenomenological model and a random-field XXZ model in the many-body localized phase. We show that the OTOC decreases in power law in a many-body localized system at the scrambling time. We also find that the OTOC can also be used to distinguish a many-body localized phase from an Anderson localized phase, while a normal correlator cannot. Furthermore, we prove an exact theorem that relates the growth of the second R\'enyi entropy in the quench dynamics to the decay of the OTOC in equilibrium. This theorem works for a generic quantum system. We discuss various implications of this theorem. 

\end{abstract}

\maketitle
\section{Introduction}
Recently, the out-of-time-order correlator (OTOC) has drawn a lot of attention in both the gravity physics, the condensed matter physics and quantum information \cite{bh1,bh2,bh3,Kitaev1,Kitaev2,prove,Yoshida,SYK1,SYK3,SYK4,Larkin,otoc1,otoc2,otoc3,otoc4,otoc5,otocexp1,otocexp2,otocexp3}. This correlator is introduced as 
\begin{equation}
F(t)=\langle \hat{W}^\dagger(t)\hat{V}^\dagger(0)\hat{W}(t)\hat{V}(0)\rangle_\beta,
\label{otoc}
\end{equation}
where $\hat{W}(t)=e^{i\hat{H}t}\hat{W}e^{-i\hat{H}t}$ and $\langle ...\rangle_\beta$ denotes averaging over a thermal ensemble at temperature $1/\beta=k_\text{B}T$. In the context of condensed matter physics, the OTOC diagnoses the chaotic behavior. The exponential deviation of the OTOC defines the Lyapunov exponent $\lambda_\text{L}$ \cite{Larkin,bh1,Kitaev1}. In the gravity context, for systems that can be described holographically by an Einstein gravity, it is shown that the Lyapunov exponent saturates $2\pi/\beta$ \cite{bh1,bh2,bh3}. More remarkably, it is shown that the Lyapunov exponent will actually be bounded by $2\pi/\beta$ \cite{prove}. It is thus conjectured that a quantum mechanical system that saturates the bound has a holographic dual to a black hole \cite{prove}. A concrete ``Sachdev-Ye-Kitaev" model \cite{SY,SYK5,Kitaev2} has been shown to display a black hole dual \cite{SYK1, SYK2, SYK3} and to have a Lyapunov exponent $\lambda_\text{L}=2\pi/\beta$ \cite{Kitaev2,SYK4}.

In this paper, we ask the question that whether the OTOC can be used beyond diagnosing the chaos and for systems that do not have (and are not even close to have) the holographic dual. For this motivation, we consider the OTOC in the many-body localized (MBL) system, which is not chaotic and even does not satisfy the Eigenstate Thermalization Hypothesis due to the existence of many local integrals of motion \cite{mbl1,mbl2,phe1,phe2,phe3}. Instead of an exponential deviation, we analytically show that the OTOC power-law decays in an MBL system. This is a clear distinction between an MBL phase and a thermalized phase. 

In the discussion of the MBL, it is often asked how to distinguish an MBL state from an Anderson localized (AL) state \cite{log growth1,log growth2,MBLAL,MBL2,MBL5}. The later is known as a non-interacting phenomenon. It is known that, after a sudden quench and following a linear growth of the entropy at the initial time, for the AL phase entropy will stay nearly as a constant, while for the MBL phase entropy will continuously grow logarithmically due to the interaction induced dephasing \cite{log growth1,log growth2,MBLAbanin,log growth3,log growth4}. Here we show that in the MBL phase the OTOC decreases during the time interval when the entropy logarithmically grows; while in the AL phase the OTOC remains as a constant. On the other hand, the normal correlators always remain constant in both the MBL phase and the AL phase. Thus we propose that the behavior of OTOC can be used to distinguish the MBL and the AL, while normal correlators cannot. 

These calculations reveal a potential connection between the growth of the entropy after a sudden quench and the decay of the OTOC. Motivated by this insight, we prove an exact theorem that builds up a rigorous connection between these two. We should emphasize that, although the insight comes from the explicit calculation in the MBL phase, the theorem holds for \textit{any} quantum system. Various implications of this theorem are also discussed.

\section{Phenomenological Model}

 A one-dimensional model with local two-state degrees of freedom was proposed as a phenomenological model for an MBL phase \cite{phe1,phe2,phe3}
\begin{equation}
\hat{H} = \sum_i h_i \hat{\tau}_i^z + \sum_{ij} J_{ij} \hat{\tau}_i^z \hat{\tau}_j^z +\ldots.  \label{H0}
\end{equation} 
where $\hat{\tau}_i$ are local Pauli operators for the ``l-bit" and denote the local integrals of motion within a localization length $\xi$. $h_i$ are random Zeeman field uniformly distributed between $[-h, h]$. $J_{ij}=\tilde{J}_{ij}\exp(-|i-j|/\xi)$ describes interaction between different l-bits, and $\tilde{J}_{ij}$ are uniformly distributed between $[-J, J]$. Each eigenstate of this Hamiltonian can be written as $|n\rangle=|\tau^z_1\tau^z_2\dots\rangle$, where $\tau^z_i=\uparrow$ or $\downarrow$. Since this model has effectively considered the physics within a localization length by a local integral of motion and the detailed process within a localization length has been ignored,  for our analysis below, the initial time $t=0$ should be interpreted in a real model as some finite time when the initial process within a localization length is completed. 

Let us consider the infinite temperature case where we can simply sum over all the states with equal weight in calculating $F(t)$. Here we choose $\hat{W} = \hat{\tau}_i^x$, $\hat{V} = \hat{\tau}_j^x$ so the OTOC is given by
\begin{equation}
F(t) = \frac{1}{2^D} \sum_n \langle n |\hat{U}^\dag \hat{\tau}_i^x \hat{U} \hat{\tau}_j^x  \hat{U}^\dag \hat{\tau}_i^x \hat{U} \hat{\tau}_j^x |n\rangle,
\end{equation}
where $\hat{U}=e^{-i\hat{H}t}$ and $D$ is the number of sites. It is straightforward to show that $\langle n |\hat{U}^\dag \hat{\tau}_i^x \hat{U} \hat{\tau}_j^x  \hat{U}^\dag \hat{\tau}_i^x \hat{U} \hat{\tau}_j^x |n\rangle= e^{\pm i4J_{ij}t}$, where $+$ ($-$) is taken when the spins on $i$- and $j$-sites are parallel (anti-parallel). Averaging over $|n\rangle$ leads to
\begin{equation}
F(t) = \cos\left( 4J_{ij}t \right). \label{Ft1}
\end{equation}
Further averaging over all random configurations results in
\begin{equation}
\overline{F}(t) = \frac{\sin (4J \exp(-|i-j|/\xi)t)}{4J \exp(-|i-j|/\xi)t}. \label{result1}
\end{equation}

Before proceeding, we would like to make a few comments on the result Eq. \ref{result1}. (i) Eq. \ref{result1} can be expanded as $1+\alpha t^2$ for the early-time behavior. The absence of linear $t$ term means that at early time the OTOC deviates from unity in power law instead of exponentially. This shows the difference in the OTOC between an MBL state and a thermalized state. When the distribution function of $\tilde{J}_{ij}$ changes or higher order terms in the Hamiltonian Eq. \ref{H0} are included, this power law behavior is quite robust while $\alpha$ is a non-universal value and will change correspondingly. (ii) $J=0$ describes the AL limit where $F(t)$ becomes a constant. This shows that the OTOC can also distinguish the MBL phase from the AL phase.  (iii) The typical time scale of the decay time is given by
\begin{equation}
t_0=\frac{\pi}{4J}e^{|i-j|/{\xi}},
\end{equation}
which increases exponentially as the the distance between $i$- and $j$-sites increases. 

\section{Random-field XXZ Model} 

We now come to a more microscopic model for MBL, that is the one-dimensional XXZ model in a random magnetic field \cite{log growth1,log growth2,mbl-transition}
\begin{equation}
\hat{H} = \sum_i J_\perp(\hat{s}_i^x\hat{s}_{i+1}^x+\hat{s}_i^y\hat{s}_{i+1}^y)+ J_z \hat{s}_i^z\hat{s}_{i+1}^z+ h_i \hat{s}_i^z. \label{XXZ}
\end{equation}
Here $\hat s^{x,y,z}_i$ are three spin operators at site-$i$, $J_\perp$ and $J_z$ are both constants, and $h_i$ are random fields uniformly distributed among $[-h,h]$. Using a Jordan-Wigner transformation to map this model into a spinless fermion model, $\hat s_i^z \hat s_{i+1}^z$ gives a nearest neighbor interaction between fermions. Thus in this model, $J_z$ represents the interaction effect. 

\begin{figure}
\includegraphics[width=0.45\textwidth]{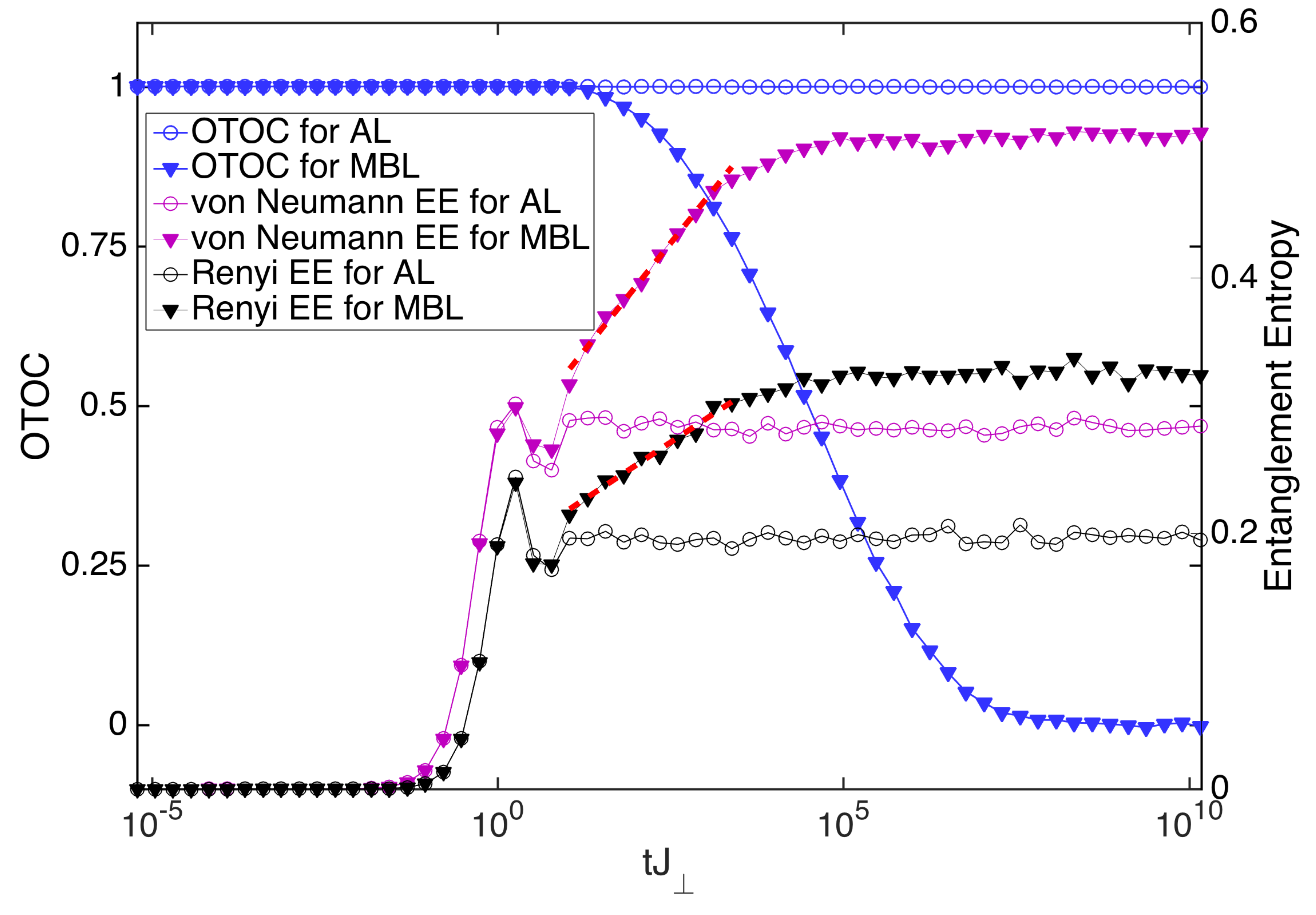}
\caption{The calculation of the von Neumann entropy, the second R\'enyi entropy and the OTOC for the MBL and the AL cases in random-field XXZ model Eq. \ref{XXZ}. The OTOC has been rescaled to drop from unity. The horizontal axis is $tJ_\perp$ in the logarithmic scale. The calculation is done for on an $8$-site model with open boundary condition, and is averaged over $10^3$ disorder configurations. Here $J_\perp>0$, $h_i/J_\perp$ is uniformly distributed between $[-5,5]$. For the MBL case $J_z/J_\perp=0.2$ where the system is known to be fully localized \cite{mbl-transition}. For the AL case $J_z=0$.  }
\label{OTOCbi}
\end{figure}

In Fig. \ref{OTOCbi} we show the von Neumann entropy, the second R\'enyi entropy (RE) and the OTOC for both the MBL case and the AL case. For the entropy calculation, the system is divided into two parts $A$ and $B$, where $A$ ($B$) is the left (right) half of this eight-site system. The second R\'enyi entropy is defined as $S^{(2)}_A=-\log \text{Tr}_A \hat{\rho}^2_A$ and $\rho_A=\text{Tr}_B\rho$. The initial state is prepared in a N\'eel state along $\hat{z}$ direction, and evolves from there under the XXZ Hamiltonian Eq. \ref{XXZ}. This initial state preparation can in fact be viewed as a global quench. For the OTOC calculation, we choose $\hat{W}$ as $\hat{s}_x$ at site $i=2$ and $\hat{V}$ as $\hat{s}_x$ at site $j=8$. The temperature is also set at infinity and we sum over all configurations with equal weight. We do check other choices of operators and most of the OTOCs all behave similarly. 

From Fig. \ref{OTOCbi} one can see that, after a linear increase at the initial time ($0<t\lesssim1/J_\perp$), both two entropys saturate for the AL case, which corresponds to the thermalization process within a localization length. Then the entropy continuously grows logarithmically only for the MBL case. The von Neumann entropy and the second R\'enyi entropy behave similarly. For the MBL case, at the time scale that entropy starts logarithmic growth, the OTOC also starts to drop. While in the AL case, the OTOC always remains constant. We also calculate the normal correlators in this model and find they always remain as constants in both the MBL phase and the AL phase. These results are consistent with the results from the phenomenological model. 

\section{The OTOC-RE Theorem}

 Motivated by the calculation above, here we prove a general theorem as 

\textbf{Theorem.} Consider a system initialized at $T=\infty$. After being quenched by an arbitrary operator $\hat{O}$ at $t=0$, we divide it into two subparts \textit{A} and \textit{B} and considering the second R\'enyi entropy $S^{(2)}_A$. The growth of this second R\'enyi entropy is related to the OTOC of the original equilibrium state via
\begin{eqnarray}
\exp(-S^{(2)}_A)=\sum_{\hat{M}\in B}\langle\hat{M}(t)\hat{V}(0)\hat{M}(t)\hat{V}(0)\rangle_{\beta=0} \label{theorem}
\end{eqnarray}
where $\hat{V}=\hat{O}\hat{O}^\dagger$ and the summation is taken over a complete set of operators $\hat{M}$ in the part \textit{B}. Here we have chosen the normalization condition for $\hat{M}$ and $\hat{O}$ as $\sum_{\hat{M}\in B}M_{ij}M_{lm}=\delta_{im}\delta_{lj}$, $\text{Tr}[\hat{O}\hat{O}^\dagger]=1$.

Before the proof of this theorem, we would like to add a few remarks on this theorem:

i) Let us first explain what quench and quench operators mean here. Suppose $\ket{\Psi_n}$ is one of the eigenstates of the Hamiltonian, here quench means suddenly apply an operator $\hat{O}$ to $\ket{\Psi_n}$. For instance, for a spin model, one can flip a spin at site-i via a quench operator $\hat{S}^{-}_i$. The wave function after the quench becomes $\hat{S}^{-}_i\ket{\Psi_n}$, which is no longer an eigenstate of the Hamiltonian and it will start to evolve and the entropy will increase. For a mixed state, the quench operator will apply identically to every state, for instance, here at infinite temperature, initially $\hat \rho \propto \hat I$, and after the quench, $\hat \rho \propto \hat{O}\hat{O}^\dag$ and the density matrix will also start to evolve and leads to increasing of entropy. For finite temperature case, initially $\hat \rho \propto e^{-\beta \hat{H}}$, and after quench it becomes $\hat \rho \propto \hat{O}e^{-\beta\hat{H}}\hat{O}^\dag$.

ii) This theorem applies to generic quantum systems, no matter whether they are chaotic, thermalized, localized or not. It is independent of how to divide the system into A and B subparts. And the quench operator $\hat{O}$ can be either a global one or a local one.

\begin{table}[t]
  \centering 
\begin{tabular}{|l|l|l|ll}
\cline{1-3}
 &  RE & OTOC  &    \\ \cline{1-3}
 T  & Linear increase \cite{cft}  &  Exponential decay  &    \\ \cline{1-3}
 MBL & Logarithmic increase \cite{das}   & Power law decay & \\ \cline{1-3}
 AL  & Constant  & Constant & \\ \cline{1-3}
\end{tabular}

\caption{A comparison of the behavior for the growth of the second R\'enyi entropy (RE) and the decay of the OTOC in the thermalized phase (denoted by ``T"), the MBL phase and the AL phase.  }\label{compare}
\end{table}

iii) This theorem builds up a general relation between the OTOC and the R\'enyi entropy. We summarize this correspondence in different phases in the Table \ref{compare} . With this relation, previous results on the second R\'enyi entropy \cite{Dong Xi,renyi1,renyi2,renyi3,renyi4,renyi5} can now be used to infer the properties of the OTOC. 

Here we should remark that another relation between the OTOC and the R\'enyi entropy has been derived in Ref. \cite{Yoshida}. The difference is that, in Ref. \cite{Yoshida}, the entropy are defined between the input and the output Hilbert space using the operator-state mapping. While in our case, both the OTOC and R\'enyi entropy are defined in the physical system. 

iv) We should note that the L.H.S. of Eq. \ref{theorem} is a quantity measured from a quenched \textit{non-equilibrium} system, while the R.H.S. is a correlator for an \textit{equilibrium} system. Thus, this theorem establishes a relation between the correlation in the equilibrium and a quantity in the dynamical process. In this sense, it shares the same spirit of the linear response theory for the normal correlators, which says the normal correlations in an equilibrium system can be related to some observables after adding a time-dependent perturbation to the Hamiltonian. Thus, to make a comparison, normal correlator measures the response of the observables to a perturbation; while the OTOC measures the response of the entropy to a quench.

v) This theorem can be generalized to the finite temperature as
\begin{equation}
\exp(-S^{(2)}_A)=\sum_{\hat{M}\in B}
\text{Tr}[\hat{M}(t)\hat{O}e^{-\beta\hat{H}}\hat{O}^\dag\hat{M}(t)\hat{O}e^{-\beta\hat{H}}\hat{O}^\dag]. \label{theorem2}
\end{equation}
One can view the R. H. S. of the Eq. \ref{theorem2} as an OTOC at $T=\infty$ with $\hat{V}=\hat{O}e^{-\beta\hat{H}}\hat{O}^\dag$, or if $\beta$ is not too large, each term in the R.H.S. of Eq.\ref{theorem2} approximates to the OTOC at $\beta^\prime=2\beta$ as $\text{Tr}[ e^{-2\beta\hat{H}}\hat{M}(t)\hat{O}\hat{O}^\dag\hat{M}(t)\hat{O}\hat{O}^\dag]$. This theorem can also be generalized to the higher order R\'enyi entropy, where $S^{(n)}_\text{A}$ will be related to a correlation function with $4n-2$ operators.

\begin{figure}[t]
\includegraphics[width=0.45\textwidth]{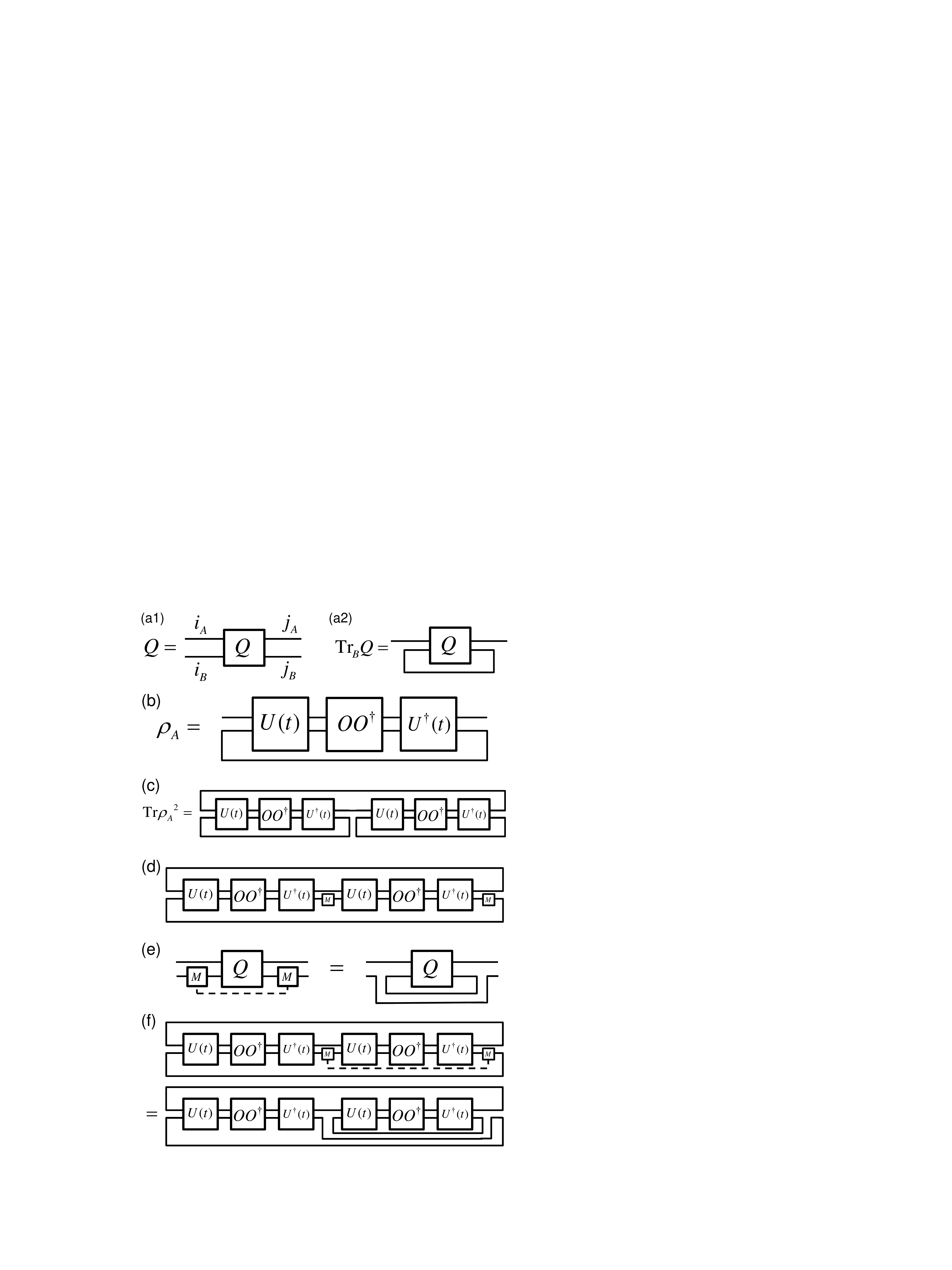}
\caption{Diagrammatic illustration of how to prove the OTOC-RE theorem. Please see the main text for details.  }
\label{diagram}
\end{figure}

Now we outline how this theorem is proved. For convenience, we first introduce a set of diagrams. For a system divided into subsystems $A$ and $B$, denote $\{|i\rangle_\text{A}\otimes|i\rangle_\text{B}\}$ as a complete set of bases in the Hilbert space, an arbitrary operator $\hat{Q}=\sum\limits_{ij}Q_{ij}|i\rangle_\text{A}\otimes|i\rangle_\text{B}\langle j|_\text{A}\otimes\langle j|_\text{B}$ is presented diagrammatically in Fig. \ref{diagram}(a1). In this representation, $\text{Tr}_\text{B}\hat{Q}$ can be described by connecting states in the subpart $B$, as presented by Fig. \ref{diagram}(a2). 

Consider a system at $T=\infty$, the initial density matrix $\hat \rho\propto \hat I$. After the quench by operator $\hat{O}$ and let the system evolve under the Hamiltonian $\hat{H}$ by time $t$, the density matrix becomes $\hat \rho=\hat{U}(t)\hat{O}\hat{O}^\dag\hat{U}^\dag(t)$. Then $\hat \rho_\text{A}$ will be represented as Fig. \ref{diagram}(b), and straightforwardly, $\text{Tr}\hat \rho_\text{A}^2=e^{-S^{(2)}_\text{A}}$ is presented by Fig. \ref{diagram}(c).

Now we consider each OTOC on the R.H.S. of Eq. \ref{theorem}, which is
\begin{align}
\text{Tr}[\hat{M}(t)\hat{V}(0)\hat{M}(t)\hat{V}(0)]&=\text{Tr}[\hat{U}^\dag\hat{M}\hat{U}\hat{V}\hat{U}^\dag\hat{M}\hat{U}\hat{V}]\nonumber\\
&=\text{Tr}[\hat{U}\hat{V}\hat{U}^\dag\hat{M}\hat{U}\hat{V}\hat{U}^\dag\hat{M}].
\end{align}
Note that $\hat{V}$ is taken as $\hat{O}\hat{O}^\dag$ and $\hat{M}$ only acts on the Hilbert space of the subsystem $B$, this is shown by Fig. \ref{diagram}(d). 

\begin{figure}[t]
\includegraphics[width=0.46\textwidth]{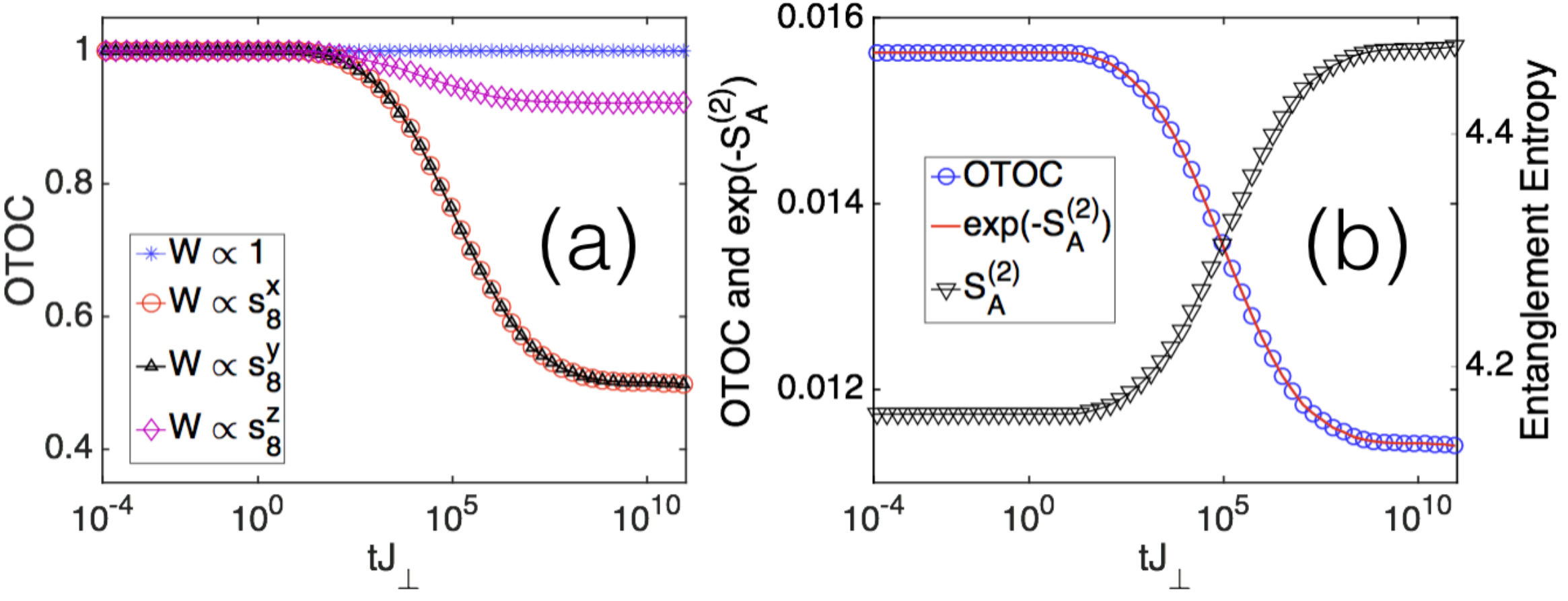}
\caption{(a) All the four OTOCs with $\hat{V}=(1+2\hat s^x_i)/2^D$ ($i=1$) and $W=1/\sqrt{2}$, $\sqrt{2}\hat s^x_j$, $\sqrt{2}\hat s^y_j$ and $\sqrt{2}\hat s^z_j$($j=8$), respectively. And all the OTOCs have been rescaled to drop from unity. (b) The summation of four (unrescaled) OTOCs, the second order R\'enyi entropy $S^{(2)}_\text{A}$ (trace out $j=8$ site which is the subsystem $B$) after quench by operator $\hat{O}=(1+2\hat s^x_i)/2^{(D+1)/2}$ and $\exp(-S^{(2)}_\text{A})$.  Here the calculation is taken on an $8$-site chain, all quantities have been averaged over $10^3$ disorder configurations. $J_z/J_\perp=0.2$ and $h_i/J_\perp$ is uniformly distributed between $[-5,5]$.  }
\label{Example}
\end{figure}

Let us again consider a general operator $\hat{Q}$, and sum over a complete set of operators in the subsystem $B$, since $\sum_{\hat{M}\in B}M_{ij}M_{lm}=\delta_{im}\delta_{lj}$, we will have $\sum_{\hat{M}\in B}\hat{M}\hat{Q}\hat{M}=\text{Tr}_\text{B}\hat{Q}\otimes \hat I$, which is shown in Fig. \ref{diagram}(e). Finally, applying this identity to $\text{Tr}[\hat{M}(t)\hat{V}(0)\hat{M}(t)\hat{V}(0)]$, the R.H.S. of Eq. \ref{theorem} is presented in Fig. \ref{diagram}(f). It is clear that the result is equivalent to Fig. \ref{diagram}(c). Hence, we prove the OTOC-EE theorem.

\section{An Example of the OTOC-RE Theorem}

 We now give a concrete example of the OTOC-RE theorem and verify it with the models mentioned above. For the phenomenological model, we consider a local quench at $i$-site by operator $\hat{O}=(1+\hat \tau^x_i)/2^{(D+1)/2}$. Then the density matrix evolves as $\hat \rho(t)=\hat U(1+\hat \tau^x_i)\hat U^\dag/2^D$. For simplicity, we only consider the $j$-site ($i\neq j$) as the $B$ subsystem. After tracing out this site, the reduced density matrix can be calculated explicitly. Finally, one obtains 
\begin{equation}
S^{(2)}_\text{A}=-\log\left(\frac{1}{2^D}\left(3+\cos(4J_{ij}t)\right)\right).
\end{equation}
This is the result before taking a disorder average. If higher order terms in Hamiltonian Eq. \ref{H0} are included, more cosine functions depending on the coefficients of higher order terms will show up as small correction.
In this case, it is also straightforward to show that $\hat{V}=\hat{O}\hat{O}^\dag=(1+\hat \tau^x_i)/2^D$ and the complete set of operators in the subsystem $B$ is $\hat \tau^0_j/\sqrt{2}$, $\hat \tau^x_j/\sqrt{2}$, $\hat \tau^y_j/\sqrt{2}$ and $\hat \tau^z_j/\sqrt{2}$. We can show that for $\hat W=\hat \tau^0_j/\sqrt{2}$ and $\hat \tau^z_j/\sqrt{2}$, the OTOC equals to $1/2^D$, while for $\hat W=\hat \tau^x_j/\sqrt{2}$ and $\hat \tau^y_j/\sqrt{2}$, the OTOC equals $(1/2^D)(1/2+\cos(4J_{ij}t)/2)$. Thus, the summation of all four OTOCs equals to $\exp(-S^{(2)}_\text{A})$.

In the random XXZ model, similarly, we consider a system quenched by operator $\hat{O}=(1+2\hat s^x_i)/2^{(D+1)/2}$, and we choose the $j$-site as the subsystem $B$. Here in our numerical calculation of eight sites, we take $i=1$ and $j=8$. To compute OTOC, we have $\hat{V}=(1+2\hat s^x_i)/2^D$, and the complete set of operators in subsystem $B$ are $1/\sqrt{2}$, $\sqrt{2}\hat s^x_j$, $\sqrt{2}\hat s^y_j$ and $\sqrt{2}\hat s^z_j$. All these four OTOCs are shown in Fig. \ref{Example}(a). One can see that except for $\hat W=1/\sqrt{2}$, all others decay. The summation of all four OTOCs is shown in Fig. \ref{Example}(b), and it is compared to the second R\'enyi entropy $S^{(2)}_\text{A}$ and $\exp(-S^{(2)}_\text{A})$. We can see that the summation of OTOC perfectly coincides with $\exp(-S^{(2)}_\text{A})$.

\section{Final Remarks} In summary, our results build up the connection between the OTOC in equilibrium and the growth of entropy after a quench. This OTOC-RE theorem will have many implications in various systems. The MBL system is explicitly discussed here with both the phenomenological model and the random XXZ model. Several MBL systems have now been realized in cold atom and trapped ion systems \cite{mbl_exp1,mbl_exp2,mbl_exp3,mbl_exp4,mbl_exp5}, and the second R\'enyi entropy has also been recently measured in the cold atom setting \cite{EE}. Our proposal can be verified in the cold atom systems. 

\textit{Acknowledgment.} We would like to thank Yingfei Gu, Chaoming Jian, Xiaoliang Qi and Xi Dong for helpful discussions. This work is supported by MOST under Grant No. 2016YFA0301600, NSFC Grant No. 11325418 and Tsinghua University Initiative Scientific Research Program.

\textit{Note Added.} When we complete this work, we become aware that the OTOC of an MBL system has also been studied by Ref. \cite{Xie_Chen}.

\end{document}